\documentclass[eqsecnum,preprint,prd,aps,nofootinbib]{revtex4}
\usepackage{epsfig}
\usepackage{amssymb}
\usepackage[dvips]{color}
\usepackage[latin1]{inputenc}
\usepackage{graphicx}

\newcommand{\nc}{\newcommand}
\nc{\be}{\begin{equation}} \nc{\ee}{\end{equation}} 
\nc{\bea}{\begin{eqnarray}} \nc{\eea}{\end{eqnarray}}
\nc{\bt}{\begin{tabular}} \nc{\et}{\end{tabular}} 
\nc{\ba}{\begin{array}} \nc{\ea}{\end{array}}
\nc{\dy}{\displaystyle} \nc{\pr}{{\rm I}} \nc{\se}{{\rm II}} 
\nc{\w}{{\rm w}} \nc{\s}{{\rm s}} \nc{\um}{{1\over
2}} \nc{\Hc}{{{\cal H}_{\rm c}}}

\def\s{\sigma}

\begin{document}
\begin{center}
\bibliographystyle{article}
{\Large \textsc{On the transition from complex to real scalar fields
in modern cosmology}}
\end{center}

\date{\today }

\author{Giampiero Esposito}
\email{giampiero.esposito@na.infn.it}
\affiliation{Istituto Nazionale di Fisica Nucleare, Sezione di Napoli,
Complesso Universitario di Monte S. Angelo, Via Cintia
Edificio 6, 80126 Napoli, Italy}
\author{Raju Roychowdhury}
\email{raju@na.infn.it}
\affiliation{Istituto Nazionale di Fisica Nucleare, 
Sezione di Napoli, Complesso
Universitario di Monte S. Angelo, Via Cintia Edificio 6, 80126 Napoli, Italy}
\affiliation{Dipartimento di Scienze Fisiche,
Complesso Universitario di Monte S. Angelo,
Via Cintia Edificio 6, 80126 Napoli, Italy}
\author{Claudio Rubano}
\email{claudio.rubano@na.infn.it}
\affiliation{Istituto Nazionale di Fisica Nucleare, Sezione di Napoli,
Complesso Universitario di Monte S. Angelo, Via Cintia
Edificio 6, 80126 Napoli, Italy}
\affiliation{Dipartimento di Scienze Fisiche, Complesso
Universitario di Monte S. Angelo, Via Cintia Edificio 6,
80126 Napoli, Italy}
\author{Paolo Scudellaro}
\email{scud@na.infn.it}
\affiliation{Istituto Nazionale di Fisica Nucleare, 
Sezione di Napoli, Complesso
Universitario di Monte S. Angelo, Via Cintia Edificio 6, 80126 Napoli, Italy}
\affiliation{Dipartimento di Scienze Fisiche,
Complesso Universitario di Monte S. Angelo,
Via Cintia Edificio 6, 80126 Napoli, Italy}

\begin{abstract}
We study some problems arising from the introduction of a complex 
scalar field in cosmology, modelling its
possible behaviors in both the inflationary and dark energy stages 
of the universe. Such examples contribute to show that,
while the complex nature of the scalar field can be indeed important 
during inflation, it loses its meaning in
the later dark--energy dominated era of cosmology, when the phase 
of the complex field is practically constant,
and there is indeed a transition from complex to real scalar 
field. In our considerations, 
the Noether symmetry approach turns out to
be a useful tool once again. We arrive eventually at a potential 
containing the sixth and fourth powers of the
scalar field, and the resulting semiclassical quantum cosmology is 
studied to gain a better understanding of the inflationary stage.
\end{abstract}

\maketitle

\section{Introduction}

The standard treatment of an inflationary scenario, as well as of 
dark energy, makes use of one or more real scalar fields, variously 
coupled to gravity and matter. On the other hand, from the point of view 
of Quantum Field Theory, such scalar fields can be taken to be
either real or complex. The main goal of this paper is to get a deeper
understanding of the transition from complex to real scalar fields, 
also contributing to implement the idea that 
the phase of a complex field characterizing the
epoch of dark energy is practically constant, so that it may be set 
to zero safely. We will not try here to
prove this result in general form, but in particular cases of wide 
interest. The general proof is postponed to
future studies.

On the other hand, during the inflationary epoch, the phase should 
very rapidly be driven to be constant, if not
already, so that the assumption of real fields would turn out to be 
justified. We shall give an example in which
this is not entirely obtained, so that further investigation of this 
point also seems to be necessary. Although there exists a large
amount of literature concerning such kind of problems
(see Refs. \cite{kasuya01,gu01,boyle02,wei05,zhao06} for example), 
here we want to look at things from a different point 
of view, adding new features and offering other examples for the
discussion. Furthermore, we do not enter at all into the domain 
of the open question whether there exists 
only a scalar field (or many of them)
driving both inflationary and dark energy epochs or not, since we 
limit ourselves just to consider the two
evolutionary stages as separated.

As a matter of fact, another important feature of this work consists 
in the application of the Noether symmetry
approach to the problem. This procedure was applied for the first time 
by some of us
\cite{deritis90,capozziello96} in the context of a minimally coupled 
real scalar field. Since then, a lot of
different results were found in various situations (as can be seen in 
Ref. \cite{capozziello96}, for instance,
just accounting for the first results only).

Here we try to find a Noether charge, for some suitable choice of the 
potential ruling one single scalar field.
The first interesting result is that, by virtue of the phase and of the 
fact that the potential depends only on
the modulus, there is always a trivial symmetry and a cyclic variable 
(the same holds also for the hessence case). 
The associated charge (below called $Q$)
was, on the other hand, already known (see Refs. 
\cite{gu01,boyle02,wei05,zhao06} and references therein).
However, the presentation which we give here seems to us more elegant 
both in terms of the procedure and the
interpretation, and we shall make use of this constant in order to make 
most of our considerations. What is more
interesting is the fact that, using the Noether symmetry approach, we 
may prove that no other symmetry is found,
except for the case $Q=0$, which indeed gives back the real case. Since 
we have proved elsewhere \cite{deritis90,capozziello96,rubano02}
that, in the real case, a class of symmetries does exist, we see that 
the more fundamental picture of a complex
field breaks the symmetry, which may thus be obtained only as a limiting 
case, when the field becomes real.

In Sec. II, we sketch the basis of our problem and find a trivial Noether 
charge. The following Sec. III is
devoted to the possibility of determining other symmetries via the 
Noether symmetry approach with an exponential
potential, while Sec. IV studies the equations of motion and the 
possibility to look at the presence of a
complex scalar field as a sort of perturbation with respect to the real 
scalar field case, hence commenting on
the numerical solution so found. This is done in the two separate 
situations of dark energy and inflation.
Section V studies the semiclassical quantum cosmology for the 
scalar-field potential obtained in Sec. IV, with
application to a better understanding of the conditions for inflation. 
Finally, in Sec. VI some conclusions are
drawn.

\section{Setting the problem and finding a trivial charge}

The observable universe and its large scale structure are usually 
considered to have their origins in an early
inflationary stage. Driven by a single scalar field, inflation in 
fact yields a distribution of adiabatic
density perturbations capable to give an account of what we observe 
today in the large scale. On the other hand,
string theory or other higher-dimensional field theories consider 
a number of scalar fields, which
could all play a role during this early stage of evolution of the 
universe. This has of course to be taken into
account when dealing with inflation and its consequences.

Even if considering only one scalar field can be considered as a 
good starting point to understand the main issues of many
fundamental problems, for sake of completeness in what follows we sketch 
something about the possibility of considering
more than one scalar field. Because of the introductory type of 
discussion we are aiming to do, however, we then
continue our treatment by using a single scalar field.

\subsection{More than one scalar field for the inflationary epoch}

Inflationary dynamics and the spectrum of primordial perturbations 
are drastically affected by considering
multiple scalar fields \cite{wands08,bassett06}. For instance, unlike 
the single-field scenario, in such a case
there appear also the nonadiabatic density perturbations, which 
could lead to detectable specific features in
future observations.

Klein--Gordon and Friedmann equations with multiple (non 
mutually interacting) scalar fields labelled by
$i$'s are written in the homogeneous and isotropic flat case as \cite{wands08}
\begin{eqnarray}
\ddot{\varphi}_i + 3H\dot{\varphi}_i + 
\frac{\partial}{\partial \varphi_i}(\sum_j U_j) &=& 0\,, \\
3H^2 - 8\pi G \left( \sum_i \frac{1}{2}{\dot{\varphi}_i}^2 
+ V \right)=0,
\end{eqnarray}
$H$ being the Hubble parameter and $U_j$ the potential of the $j$-th 
scalar field. Here, we are also assuming
that the global potential energy is given by the sum of such 
single potentials, as
\begin{equation}
V \equiv \sum_j U_j\,.
\end{equation}
Since the Hubble expansion is affected by the sum of all single 
potentials, it is thus clear that the overall
field dynamics can be substantially changed in the presence of 
multiple scalar fields, even if the single
potentials are not affected. There are, of course, different ways to 
consider many fields in inflation and we
will not touch upon them here. See Ref. \cite{wands08} for further 
comments and citations to literature.

Even if there is the possibility to reduce the background dynamics to an 
equivalent single field with only a
potential \cite{wands08,malik99}, 
it is nevertheless important to stress again 
that there are meaningful differences
between the scenario with multiple fields and the one with a single 
scalar field. While the evolution in the latter
case can be shown to be independent of the initial conditions, the 
inflationary dynamics in the former may
not. But, while keeping some distinct features of a multiple field 
scenario, it can anyway lead to observable
predictions which may not depend on the initial conditions, as 
in Nflation \cite{lyth99,kim06}.

Implementing the inflationary paradigm with many scalar fields 
has gained an increasing interest, since we are
now facing a period of greater precision in cosmological observations, 
which allows the future possibility to
discriminate between these various forms of inflationary stages. 
But, for what concerns us here, we shall return
to the single field model in the following.

\subsection{A single scalar field}

It is easy to verify that the Friedmann and Klein--Gordon equations for 
a spatially flat FLRW metric and a
complex scalar field $\phi$ may be derived from the following 
point Lagrangian
\begin{equation}\label{1}
L=3a\dot{a}^{2}-\frac{a^{3}}{2}(\dot{\varphi}_{1}^{2}
+\dot{\varphi}_{2}^{2})+a^{3}V(\varphi)\,,
\end{equation}
where we have used the metric signature $(+\,-\,-\,-)$ and 
adopted the units in which $8\pi G = 1$, while $\varphi_{1,2}$ are the
real and imaginary parts of the field $\phi$ and 
$\varphi \equiv \sqrt{\varphi_{1}^{2}+\varphi_{2}^{2}}$ is its modulus,
so that one can write $\varphi_{1}=\varphi \cos \theta$ and 
$\varphi_{2}=\varphi \sin \theta$.

It is important to point out that the Lagrangian is the same in  
absence of matter, as well as, up to
a constant, in its presence in the form of dust. 
The difference consists only in the value of the
Energy Function associated, as usual, with $L$
\begin{equation}\label{2}
E_{L} \equiv
\sum_{i}\frac{\partial L}{\partial\dot{q}_{i}}\dot{q}_{i}-L\,,
\end{equation}
which of course, in our case, should not be interpreted as the 
physical energy. In the case of vacuum $E_{L}$
vanishes, while in the case of dust $E_{L}=const.$, a value to be 
related to the present amount of matter, i.e.
$\Omega_{m0}$.

The following arguments will be therefore applied in the two separate 
cases of inflation and dark energy, with
suitable specifications when necessary. As already said, in fact, we do 
not want to enter into the realm of the
unsolved problem of a unique scalar field, describing both the primeval 
era and the present accelerated
expansion of the universe. Any relationship between the two fields 
present in such different cosmological stages is taken
as inessential for our discussion, and we simply limit ourselves to 
consider the predominance of a complex
scalar field in each one of the two epochs.

For our purposes, it is much better to introduce the phase 
$\theta$ so that the Lagrangian becomes
\begin{equation}\label{3}
L=3a\dot{a}^{2}-\frac{a^{3}}{2}(\dot{\varphi}^{2}
+\varphi^{2}\dot{\theta}^{2})+a^{3}V(\varphi)\,,
\end{equation}
and we see at once that $\theta$ is cyclic. Thus, we immediately 
obtain the conserved charge
\begin{equation}\label{4}
\Sigma\equiv -a^{3}\varphi^{2}\dot{\theta}\,,
\end{equation}
as expected. The minus sign is irrelevant, as $\Sigma$ will always 
appear in squared form, and we shall set $Q
\equiv \Sigma^{2}$. It is now clear what was stated in the introduction 
above: $\Sigma=0$ corresponds to the
real case, for which we already have a large set of interesting 
results \cite{deritis90,capozziello96,rubano02}.

In order to look for other symmetries, we need to make use of the 
charge obtained and reduce the degrees of
freedom of the system. The procedure is exactly the same as in the 
case of a central potential in classical
mechanics. First, we write the Energy Function
\begin{equation}\label{5}
E_{L}=3a\dot{a}^{2}-\frac{a^{3}}{2}(\dot{\varphi}^{2}
+\varphi^{2}\dot{\theta }^{2})-a^{3}V(\varphi)\,,
\end{equation}
and substitute for $\dot{\theta}$ to obtain
\begin{equation}\label{6}
E_{L}=3a\dot{a}^{2}-\frac{a^{3}}{2}\dot{\varphi}^{2}
-\frac{Q}{2a^{3} \varphi^{2}}-a^{3}V(\varphi)\,.
\end{equation}
Finally we have the reduced Lagrangian
\begin{equation}\label{7}
L^{\prime}=3a\dot{a}^{2}-\frac{a^{3}}{2}\dot{\varphi}^{2}+a^{3}\left(
\frac{Q}{2a^{6}\varphi^{2}}+V(\varphi)\right)\,.
\end{equation}
It should be stressed that this expression for the Lagrangian is 
\emph{not} the same as the one obtained by
direct substitution of Eq. (2.7) 
into $L$ in Eq. (\ref{3}). We have in fact reduced 
the configuration space of the problem
from $3$ degrees of freedom ($\{a,\varphi,\theta\}$) to $2$ 
($\{a,\varphi \}$), as usual in this approach. We
also observe that the system is now ``equivalent'' to that of a 
real field, endowed with an effective potential
\begin{equation}\label{8}
W(a,\varphi)\equiv \frac{Q}{2a^{6}\varphi^{2}}+V(\varphi)\,,
\end{equation}
as already pointed out elsewhere (see Ref. \cite{zhao06}, for example). 
The presence of the scale factor in $W$
substantially modifies the situation, and the field is now in a sense 
nonminimally coupled to gravity, although
in a way which is not standard. This is of course due to the fact that 
the new Lagrangian is not physical, but
obtained as a result of a reduction procedure.

\section{Looking for new symmetries}

Let us now apply the standard procedure of the Noether symmetry 
approach in order to establish nontrivial
symmetries of $L^{\prime}$. We look for a vector field on the 
reduced configuration space, properly lifted to
the tangent space \cite{abraham78,marmo85,morandi90,deritis90,capozziello96}
\begin{equation}\label{9}
X=\alpha(a,\varphi)\frac{\partial}{\partial a}
+\beta(a,\varphi)\frac{\partial}{\partial\varphi}
+\dot{\alpha}(a,\varphi)\frac{\partial}{\partial\dot{a}
}+\dot{\beta}(a,\varphi)\frac{\partial}{\partial\dot{\varphi}}\,,
\end{equation}
such that the Lie derivative $\mathcal{L}_{X}L^{\prime}$ vanishes. 
Here, the two functions $\alpha(a,\varphi)$ and $\beta(a,\varphi)$
are unknown. By virtue of the linear dependence of $X$ on
``velocities'' and the quadratic ones of $L^{\prime}$, a quadratic 
polynomial in $\dot{a}$, $\dot{\varphi}$ is
obtained. Since it must be identically zero, a set of four 
equations is finally derived, i.e.
\begin{eqnarray}
\alpha+2a\frac{\partial\alpha}{\partial a} &=& 0\,, \\
3\alpha+2a\frac{\partial\beta}{\partial\varphi} &=& 0\,, \\
6\frac{\partial\alpha}{\partial\varphi}-a^{2}
\frac{\partial\beta}{\partial a} &=& 0\,, \\
3\alpha V(\varphi)+a\beta\frac{dV}{d\varphi} &=& Q\left(  \frac{3\alpha
}{2a^{6}\varphi^{2}}+\frac{\beta}{a^{5}\varphi^{3}}\right)\,.
\end{eqnarray}
The first three equations are identical as in Refs. 
\cite{deritis90,capozziello96}, while the last one differs
because of the presence of an effective potential, by 
virtue of the supposed complex nature of the scalar field.
Thus, proceeding along the very same lines, it is possible to 
find a general solution for $\alpha$ and $\beta$,
which however, when inserted into the last equation, turns out 
to be incompatible, unless $Q=0$ (in the latter situation, of
course, a specified class of potentials is found
\cite{deritis90,capozziello96}).

It is thus proved that, in the fully complex case, no nontrivial 
symmetry exists. It should be stressed that
the potential has been simply viewed as a function of 
$\varphi \equiv \sqrt{\varphi_{1}^{2}+\varphi_{2}^{2}}$, but
the same thing can be proved also while using a potential depending on 
$\sqrt{\varphi_{1}^{2}-\varphi_{2}^{2}}$, 
so covering the well known hessence case \cite{wei05}, for instance.
As a matter of fact, in the latter situation one has
$\varphi \equiv \sqrt{\varphi_{1}^{2}-\varphi_{2}^{2}}$, where
$\varphi_{1} \equiv \varphi \cosh \theta$ and 
$\varphi_{2} \equiv \varphi \sinh \theta$, so that
$\coth \theta={\varphi_{1}\over \varphi_{2}}$. One then obtains
the same set of Eqs. (3.2)--(3.4), while Eq. (3.5) is replaced by
$$
3 \alpha V(\varphi)+a \beta {dV \over d\varphi}
=-Q \left({9 \alpha \over 2 a^{6}\varphi^{2}}
+{\beta \over a^{5}\varphi^{3}}\right).
$$
Of course, this does not change the qualitative features of our 
previous results obtained from the Noether symmetry approach.
In the case $Q=0$, instead, we recover the real scalar 
field situation, where it is well known that a class of
exponential potentials does the job \cite{deritis90,capozziello96}. 
In particular, we shall now make use of the potential
\begin{equation}\label{10}
V=V_{0}e^{-\sqrt{\frac{3}{2}}\varphi}\,,
\end{equation}
finding indeed a peculiar situation \cite{rubano02,rubano04}.

We are in fact used to see that in Quantum Field Theory some 
symmetries are generally found at a more
fundamental level, which are subsequently broken during the 
evolution. Here, on the contrary, assuming a
potential of the form in Eq. (\ref{10}), we see that a complex 
field shows initially no symmetry. Since
$Q=a^{6}\varphi^{4} \dot{\theta}^{2}$, we have that, along with 
the growth of the scale factor with time, and
provided that $\varphi$ does not contemporarily go to zero very 
rapidly, $\dot{\theta }$ should approach $0$.
The term with the initially nonvanishing $Q$ then becomes negligible 
and, at least with very good approximation,
the Noether symmetry is recovered, in contrast with the usual 
expectation. 

\section{Equations of motion and numerical perturbations}

It is now clear that $Q$ can be regarded as a parameter, 
characterizing a perturbation term in $L^{\prime}$. In
this way the real scalar field corresponds to the \emph{unperturbed} 
solution, while the phase of the complex
scalar field gives rise to the \emph{perturbed} solution. This point of 
view turns out to be particularly useful
once an exact solution for the real scalar field is available. This being 
our case, let us first write, for the reduced Lagrangian (2.10),
the Euler--Lagrange equations
\begin{eqnarray}
2\frac{\ddot{a}}{a}+\left(  \frac{\dot{a}}{a}\right)  ^{2}+\frac{1}{2}\left(
\dot{\varphi}^{2}+\frac{Q}{a^{6}\varphi^{2}}\right)  -V(\varphi) &=& 0\,, \\
\ddot{\varphi}+3\frac{\dot{a}}{a}\dot{\varphi}
+\frac{dV}{d\varphi}-\frac {Q}{a^{6}\varphi^{3}} &=& 0\,,
\end{eqnarray}
and the conservation of the Energy Function
\begin{equation}\label{11}
3a\dot{a}^{2}-\frac{a^{3}}{2}\dot{\varphi}^{2}
-\frac{Q}{2a^{3}\varphi^{2}}-a^{3}V(\varphi)=E_{L}\,,
\end{equation}
where, as we said, $E_{L}=0$ corresponds to the vacuum case.
In this case, an important consistency check is as follows.
Equation (4.1), jointly with $E_{L}=0$, yields
$$
{\dot H}=-{{\dot \varphi}^{2}\over 2}-{Q\over 2a^{6}\varphi^{2}}
$$
for the time derivative of the Hubble parameter, once that $E_{L}=0$
is used to re-express $3H^{2}$. On the other hand, $E_{L}=0$ should
have vanishing time derivative (this corresponds to preservation in
time of the Hamiltonian constraint), and this yields again, upon
expressing $\ddot \varphi$ from the Klein--Gordon equation (4.2), the
formula for ${\dot H}$ written above, since the additional contributions
of ${{\dot \varphi}Q \over a^{6}\varphi^{3}}$ occur with opposite signs
and hence cancel each other exactly.

In what follows we shall first consider, due to its brevity, the dark 
energy situation and then the inflationary
one. This gives no problem, since one has always to remember that we 
have not assumed here any relationship
between the two scalar fields driving the two different stages of the 
cosmological evolution.

\subsection{Perturbation of a dark energy solution}

Let us, now, first consider an exact solution for the dark energy 
scenario. In the case $Q=0$, as said above,
the exponential potential in Eq. (\ref{10}) exhibits a Noether symmetry. 
This allows exact integration of
equations, as shown in Refs. \cite{rubano02,rubano04}. It is also possible 
to show that the solution fits very well
with the SNIa data, as well as other tests \cite{rubano02,rubano04}.

Following the procedure of Ref. \cite{rubano04} we find it useful to 
set the unit of time to the age $t_{0}$ of
the universe. This greatly simplifies notations and, above all, is of 
great help in numerical treatments, which
are necessary when we go to the perturbed system. In these units $H_{0}$ 
turns out to be of order $1$, and it is
in fact convenient to set it exactly to $1$. 
Actually, this is not the best fit 
value, but very near, so that it can be
considered appropriate for our purposes. Other choices of the 
integration constants lead to
$a_{0}=a(t_{0}=1)=1$, and $a(0)=0$, which are quite standard. The last 
constant fixes the present value of the
scalar field $\varphi_{0}$. In the unperturbed system, 
however, this is quite arbitrary, since we are not able to observe
it in any way. Here, lies a problem: if, during the evolution of the 
unperturbed solution, one gets $\varphi=0$,
the perturbing term diverges! In order to avoid this pathology, we 
have set $\varphi_{0} \equiv \sqrt{2/3}\log \,2$,
which is not the same as in Ref. \cite{rubano04}, but appropriately shifted.

In light of all these assumptions, the solution takes the very simple form
\begin{eqnarray}
a^{3}(t) &=& \frac{t^{2}(1+t^{2})}{2}\,, \\
\varphi(t) &=& \sqrt{\frac{2}{3}}\log(1+t^{2})\,, \\
H(t) &=& \frac{2(1+2t^{2})}{3t(1+t^{2})}\,, \\
V(\varphi) &=& 4\exp(-\sqrt{3/2}\varphi)=\frac{4}{(1+t^{2})}\,, \\
\Omega_{m} &=& \frac{(1+t^{2})}{(1+2t^{2})^{2}}\,, \\
\Omega_{m0} &=& \frac{2}{9}\approx0.22\,,
\end{eqnarray}
where the value of $\Omega_{m0}$ obtained here is not the best fit one, 
by virtue of the simplified choice of
$H_{0}$. It is easy to check that the equations of motion are satisfied 
by this solution if $Q=0$.

\begin{figure}
\includegraphics{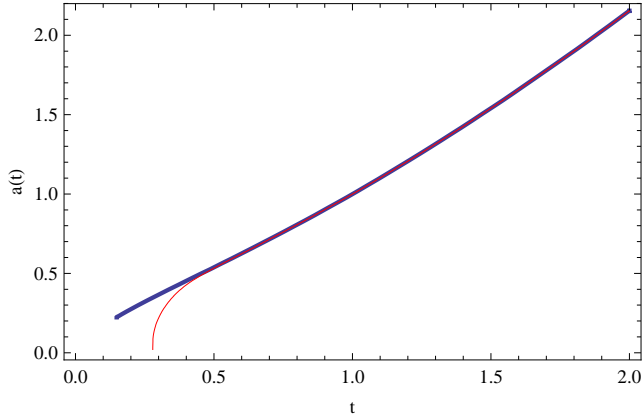}
\caption{Comparison between the exact (blue colored) and the perturbed
(red colored) scale factors $a(t)$ for
the exponential potential given in Eq. (\ref{10}),
in the dark-energy regime.} \label{f1}
\end{figure}

With our choices, at $t_{0}$ all terms in the equations are of order 
$1$, so that a small perturbation is
obtained for $Q<<1$. Introducing this term now requires a numerical 
integration, which is performed starting
from $t_{0}=1$ and integrating backwards. The choice of $Q$ is of course 
arbitrary; let us thus set $Q=10^{-4}$,
for example. The plot in Fig. (\ref{f1}) shows a comparison between the 
exact solution for the scale factor
$a(t)$ and the perturbed one in the case of the exponential potential above.

It is clearly seen that the agreement is quite good for $t\gtrsim0.3$, 
as may be checked also by simple
computation. Before this time, something dramatic happens and a singularity 
is obtained for $t\approx$ $0.088$.
This is of course due to the reduced value of the denominator in the 
perturbation term, which prevents this from
being small. A computation of the redshift at the time when the discrepancy 
becomes visible gives $z\approx1.0$,
which is of course untenable. Lowering the value of $Q$, would clearly 
push the discrepancies away, possibly up
to an epoch when the model is not applicable. But this is in fact what we 
wanted to show. {\it This dark energy model
is viable only if the perturbing term is effectively negligible}. It is 
also clear that any unperturbed system,
for which the term $a^{3}\varphi^{2}$ tends to zero in the past, will give 
similar results. Thus, a better 
exploration of this possibility is needed.

\subsection{Perturbing an inflationary solution}

Let us now move to a totally different situation, that is the earlier 
inflationary scenario. We want to see what
happens if the perturbation term is applied to an exact solution 
describing primeval inflation. The general
exact solution for the exponential potential in Eq. (\ref{10}) can be 
easily specified to the new situation,
but, unfortunately, what is then obtained is a power-law inflation, with 
behavior $a(t)\propto t^{4/3}$ \cite{capozziello96}. As is
well known, the spectrum of primordial perturbations for power-law 
inflation $a(t)\propto t^{p}$ can be
computed, and the spectral index is given approximately by
\begin{equation}\label{12}
n_s \approx 1-\frac{2}{p}\,,
\end{equation}
so that the higher is the exponent $p$, the nearer we are to the 
spectral index of the Harrison--Zel'dovich spectrum \cite{har70,zel72}.
This means that $p$ must be sufficiently large if we want 
to obtain $n_s\approx1$, in agreement with WMAP observations.

Thus, for the purpose of describing a more realistic situation, it is 
necessary to give up the Noether symmetry,
but we can anyway use an exact solution for 
the unperturbed system. It was shown long
ago \cite{lucchin85} (but see also Ref. \cite{rubano02}) 
that a power-law behavior can be obtained as a 
particular solution from the potential
\begin{equation}\label{13}
V=V_{0}\exp\left(  -\sqrt{\frac{2}{p}}\varphi\right)\,,
\end{equation}
so that
\begin{equation}\label{14}
a=t^{p}\quad;\quad\varphi=\sqrt{2p}\log\left(
t\sqrt{\frac{V_{0}}{p(3p-1)}}\right)\,,
\end{equation}
where we note that the value of $V_{0}$ only fixes a shift in $\varphi$, 
which is on the other hand important in
order to avoid singularities as above.

In this new situation, we have to fix again a suitable unit of time. 
A natural choice is to set the beginning of
inflation at $t_{{\rm in}}=1$, i.e., if we accept the usual estimate 
$t_{{\rm in}}=t_{Pl}$ (the Planck time), we are
restoring the usual Planck units. Let us also set $p=50$, which gives 
a reasonable $n$, and provides, for
$t_{{\rm end}}=3.5$, an acceptable e-folding number 
$N_{e}\approx62$. The small interval spanned by time is then good
for the numerical integration. The last constant to be fixed is 
$V_{0}$, which we can arbitrarily set to $1$.

It is again easy to find that, by setting $E_{L}=0$ into Eqs. 
(\ref{5}) and (\ref{6}), and inserting Eqs.
(\ref{13}) and (\ref{14}) into them, they are in fact verified. Now, 
in order to introduce the perturbation, it
is necessary to establish for which value of $Q$ it turns out to be 
sufficiently ``small''. A check of the terms in the
equations at the initial point $t_{{\rm in}}$ shows that now they are far 
from being of order $1$. For instance, the
modified Klein--Gordon equation gives at the initial point 
$\ddot{\varphi }=-10-0.0000112871Q$. This means that
a value of, say, $Q=10^{5}$ provides a small, but not entirely 
negligible, perturbation.

\begin{figure}
\includegraphics{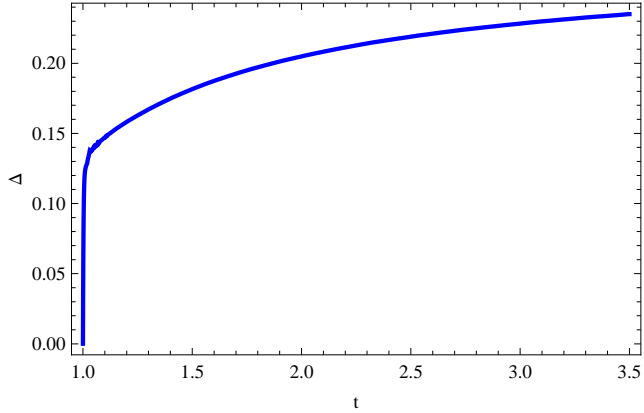}
\caption{Plot of the percentage difference $\Delta (t)$ between
the exact and perturbed solutions for the exponential potential
in the inflationary regime.} \label{f2}
\end{figure}

On performing a numerical integration from 
$t_{{\rm in}}=1$ to $t_{{\rm end}}=3.5$ and plotting the 
percentage difference $\Delta$ between the exact and perturbed 
solutions with respect to time $t$, we obtain what is plotted 
in Fig. (\ref{f2}).

This shows that such percentage difference is 
\textit{growing}, even if slowly, up 
to $0.2\%$. This is clearly small, but we
have to remember that we started from a small perturbation. 
As a matter of fact, taking a 
larger value like $Q=10^{7}$ allows the
difference to grow up to about $15\%$.

On the basis of this result we might conclude that the treatment of 
inflation by means of a real scalar field
would be a rather crude approximation. In fact, this is not the case, 
because what is really important is to
check that the phase of the scalar field is nearly constant at the 
end of inflation, even starting from a different situation.

In order to do that, let us revert to Eq. (\ref{4}), from which we get
\begin{equation}\label{15}
\dot{\theta}=-\frac{\Sigma}{a^{3}\varphi^{2}}\,,
\end{equation}
such that
\begin{equation}\label{16}
\theta=-\int\frac{\Sigma dt}{a^{3}\varphi^{2}}+C,
\end{equation}
where $C$ is an arbitrary constant.

Let us note here a subtle point. After examining the behavior of 
$\delta\equiv \dot{\theta}/\theta$, we see that
it is not dimensionless, which would force us to multiply it by an 
appropriate time scale. On the other hand,
with our units, this time scale is actually of order $1$, so that 
the result we would then get is numerically
the same as the one already deduced.

\begin{figure}
\includegraphics{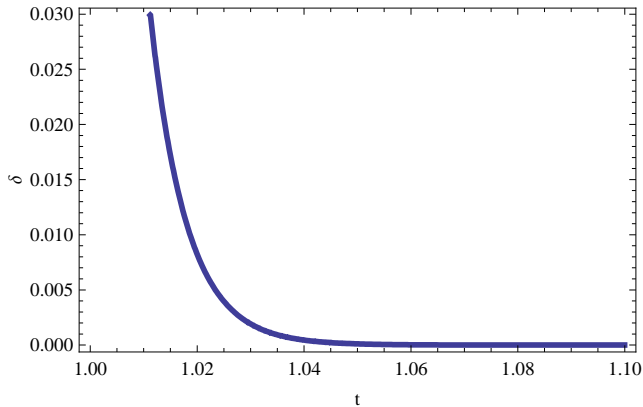}
\caption{Behaviour of $\delta(t)$.} \label{f3}
\end{figure}

Another point is more delicate. To comment on it, we shall use the 
exact unperturbed solution in the following
computation. This is justified by the fact that, as just shown 
(Fig. (\ref{f2})), it is only slightly different
from the perturbed one. Surprisingly enough, the integral 
in Eq. (4.14) can be performed in terms of a special function, i.e.
\begin{equation}\label{17}
\theta(t)=\sum \frac{\mu t+(3p-1)(\mu t)^{3p}\log(\mu t)
{\rm Ei}{((3p-1)\log(\mu t))}}{2\mu pt^{3p}\log(\mu t)}+C,
\end{equation}
where $\Sigma$ now indicates summation and $\mu\equiv \sqrt{1/(p(3p-1))}$, 
while ${\rm Ei}{(x)}\equiv
-\int_{-x}^{\infty}\frac{\exp(-t)}{t}dt$ is the exponential integral 
function. In order to determine the
integration constant $C$, 
it is possible to set the initial phase to $1$, 
for instance. This means that the
imaginary and real parts of the field are of the same order, or even 
larger, without substantial changes. By
introducing the above numbers for the parameters, we get 
$\delta (t_{{\rm in}}=1)\approx10^{-1}$ and
$\delta(t_{{\rm end}}=3.5)\approx10^{-82}$, 
showing a dramatic fall--off. The situation is illustrated by the behavior
of $\delta(t)$ shown in Fig. (\ref{f3}).

\subsection{Toy model}

\begin{figure}
\includegraphics{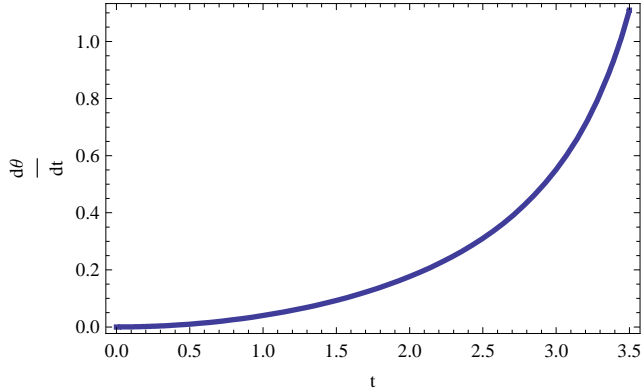}
\caption{Behavior of $\dot{\theta} = d \theta /dt$ as a
function of time $t$.} \label{f4}
\end{figure}

In this subsection, we are aiming to show that the situation can be
even more complicated, with some subtleties which do not appear in what
depicted above. For this purpose we present a toy model, which 
has the advantage of being simple and suitable
for illustrating our claim. Thus, 
let us consider the unperturbed equations and assume
that the evolution of $\varphi$ is given by $\varphi=t^{-1}$. 
From the well known relation
\begin{equation}\label{18}
\dot{\varphi}^{2}=-2\dot{H}\,,
\end{equation}
it is possible to derive
\begin{equation}\label{19}
a=\exp \left( -\frac{1}{12t^{2}} \right) \,,
\end{equation}
and it is also possible to arrive at the potential
\begin{equation}\label{20}
V=\frac{1}{12}\varphi^{6}-\frac{1}{2}\varphi^{4}\,.
\end{equation}
It is then easy to check that $V$ in (4.18) is again a special 
exact solution. We have now
\begin{equation}\label{21}
\lim_{t\rightarrow\infty}(a^{3}\varphi^{2})=0\,,
\end{equation}
so that $\dot{\theta}$ goes to infinity with time in this case
(see Fig. \ref{f4}). As a matter of fact, it appears to be divergent
already at the time $t=3.5$, as shown in Fig. 4.

The expression of Eq. (4.17) is clearly inflationary for small $t$,
because
\begin{equation}
{\ddot a}=e^{-{1\over 12 t^{2}}} {(1-18 t^{2})\over 36 t^{6}},
\end{equation}
so that inflation actually ends at $t_{f}=\sqrt{18}$. Here, it 
has been allowed to start from the very
beginning, in such a way that, 
the time scale being arbitrary, there is no
problem with the e-folding number. On the other hand, our formula
for the scale factor leads to
$$
\lim_{t \to \infty}a(t)=1,
$$
which is clearly unphysical, but we already acknowledged that it is
only a toy model, which should be anyway taken more seriously into
account only until the end of inflation. 

Let us now consider the perturbed system. On going
back to Eq. (4.1) we see that, for very small $a$, the term containing
$Q$ can dominate and there is no inflation. In order to verify this,
let us take as ``initial'' time $t_{f}=\sqrt{18}$, i.e. the time 
corresponding to the end of inflation, and let us integrate the 
Euler--Lagrange equations backwards in time. By assuming $Q=10^{-8}$,
we obtain the plot of Fig. 5 for $\ddot a$.

\begin{figure}
\includegraphics{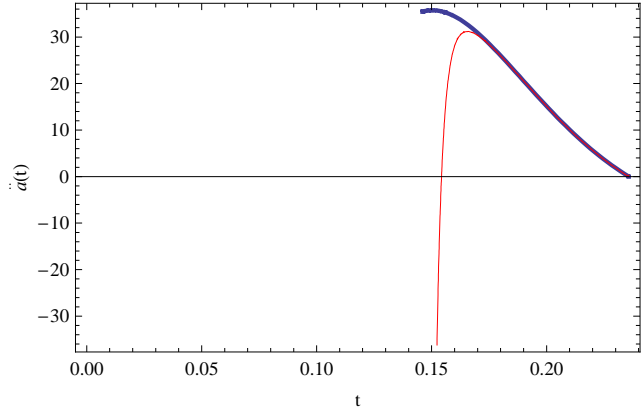}
\caption{Plot of the time evolution of $\ddot a$, for both the exact
(blue colored) and perturbed (red colored) scale factors $a(t)$,
when the exponential potential given in Eq. (3.6) is used
in the inflationary regime.}
\label{f5}
\end{figure}

We see that indeed now one has a natural way for the beginning of
inflation. This sounds interesting, but it is also clear that the
duration is now too much shortened, and the e-folding turns out
to be of order $1$! Of course, it is possible to improve the situation
by taking a much smaller value for $Q$, so small that it is impossible
to perform a numerical integration. This is a sort of fine tuning
on the phase which should be investigated further. Eventually, we
now go beyond the end of inflation, to the future of $t_{f}$. In Fig. 6
a double surprise is now obtained. First, we obtain a new limited
period of accelerated expansion. This suggests, confirming in part
what was found before in Sec. IV, that a complex scalar
field might emulate dark energy, but of course this statement should be
supported with much deeper investigation. Second, we obtain a singularity
at $t \approx 37$. It is not clear whether this is due to 
numerical integration problems. Of course, a choice of a much smaller
value for $Q$ would push these features to the right. But this is even
more interesting because, by virtue of the time scale here adopted,
this is just what is needed.

\begin{figure}
\includegraphics{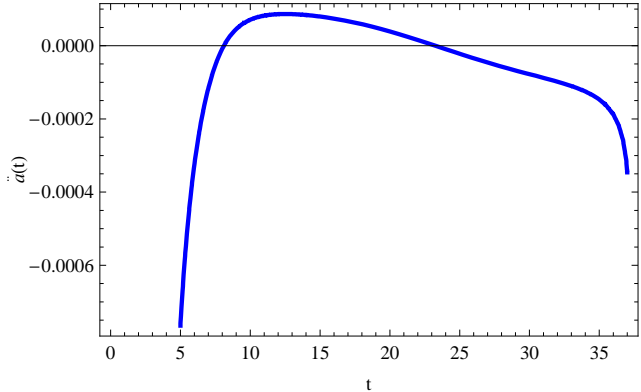}
\caption{Plot of the time evolution of $\ddot a$, showing a short
period of accelerated expansion after the end of inflation, with a
possible singularity at a certain time.}
\label{f6}
\end{figure}

\section{Semiclassical quantum cosmology with the potential of the form
$V(\varphi)=\frac{1}{12}\varphi^{6}-\frac{1}{2}\varphi^{4}$}

Since we have just found a strong motivation for studying the somewhat 
unusual potential in Eq. (\ref{20}), we
now take it again into account from a complementary point of view, i.e., 
the semiclassical approximation in
quantum cosmology. For the basic references on this vast subject, and 
for a detailed discussion of the main
issues, we refer the reader to Refs. \cite{HH83,H84,Espo88,Hall09} 
and to the many references therein.

In a minisuperspace approach to quantum cosmology, one may start from 
the point Lagrangian ruling the FLRW
models, which, in dimensionless units and with the metric signature 
$(-\,+\,+\,+)$, reads as (to be consistent with all previous sections,
our gravitational Lagrangian is $-6$ times the one in Ref.
\cite{Espo88})
\begin{equation}
L=-3a{\dot a}^{2}+3ka+{1\over 2}a^{3} 
{\dot \varphi}^{2}-a^{3}V(\varphi)\,, 
\label{(5.1)}
\end{equation}
where $k$ is positive, vanishing, negative,  
for closed, spatially flat and open universe models, 
respectively. The resulting classical Hamiltonian takes the form
\begin{equation}
H=-{1\over 12a}p_{a}^{2}+{1\over 2a^{3}}p_{\varphi}^{2} 
-3ka+a^{3}V(\varphi)\,, 
\label{(5.2)}
\end{equation}
where the conjugate momenta $p_{a}$ and $p_{\varphi}$ are given by
\begin{equation}
p_{a} \equiv {\partial L \over \partial {\dot a}} 
=-6a \dot a\,, \; p_{\varphi} \equiv {\partial L \over \partial
{\dot \varphi}} =a^{3}{\dot \varphi}\,. 
\label{(5.3)}
\end{equation}
Now we use the hat symbol for the momentum and Hamiltonian operators 
of the quantum theory, and we adopt the
following operator-ordering prescription \cite{Espo88}:
\begin{equation}
{\hat p}_{a}^{2} \equiv -{1\over a}{\partial \over \partial a} 
\left(a {\partial \over \partial a} \right)
=-{\partial^{2}\over \partial a^{2}} -{1\over a}{\partial 
\over \partial a}\,, 
\label{(5.4)}
\end{equation}
\begin{equation}
{\hat p}_{\varphi}^{2}=-{\partial^{2}\over \partial \varphi^{2}}\,, 
\label{(5.5)}
\end{equation}
where, on defining $\alpha \equiv \log(a)$, one has
\begin{equation}
{\partial \over \partial a}=e^{-\alpha}{\partial \over \partial \alpha}\,, 
\label{(5.6)}
\end{equation}
\begin{equation}
{\partial^{2}\over \partial a^{2}}=e^{-3 \alpha} 
\left({\partial^{2}\over \partial \alpha^{2}} -{\partial \over
\partial \alpha}\right)\,, 
\label{(5.7)}
\end{equation}
and hence
\begin{equation}
{\hat H}={1\over 2}e^{-3 \alpha} \left({1\over 6} {\partial^{2}\over 
\partial \alpha^{2}} -{\partial^{2}\over \partial
\varphi^{2}}\right) -3ke^{\alpha}+e^{3\alpha}V(\varphi)\,. 
\label{(5.8)}
\end{equation}
This operator can be written in a more symmetrical form by defining
$\varphi \equiv \sqrt{6}\phi$, which yields a quantum cosmological
wave function of the minisuperspace model depending only on the
$\alpha,\phi$ variables and ruled by the Wheeler--DeWitt equation
\begin{equation}
\left[{\partial^{2}\over \partial \alpha^{2}} 
-{\partial^{2}\over \partial \phi^{2}} 
+12V(\sqrt{6}\phi)e^{6\alpha}
-36ke^{4\alpha} \right]\psi(\alpha,\phi)=0\,. 
\label{(5.9)}
\end{equation}
In particular, in the spatially flat case, 
for which $k=0$, the phase $S$ of the JWKB 
ansatz $\psi(\alpha,\phi) \sim C e^{iS}$
should obey the Hamilton--Jacobi equation
\begin{equation}
\left({\partial S \over \partial \phi}\right)^{2} 
-\left({\partial S \over \partial \alpha}\right)^{2} =-12
V(\sqrt{6}\phi)e^{6\alpha}\,. 
\label{(5.10)}
\end{equation}
The comparison with the closed FLRW model studied in Ref. 
\cite{Espo88} suggests considering a factorized ansatz for $S$, i.e.
\begin{equation}
S(\alpha,\phi)=S_{1}(\alpha)S_{2}(\phi)\,. 
\label{(5.11)}
\end{equation}
Its insertion into Eq. (5.10) yields
\begin{equation}
\left({1\over S_{2}}{dS_{2}\over d\phi}\right)^{2} 
-\left({1\over S_{1}}{dS_{1}\over d\alpha}\right)^{2}
=-{12V(\sqrt{6}\phi)\over S_{2}^{2}(\phi)} {e^{6\alpha}
\over S_{1}^{2}(\alpha)}, 
\label{(5.12)}
\end{equation}
which may be solved approximately by setting
\begin{equation}
S_{1}(\alpha)=-{e^{3\alpha}\over 3}\,, \; S_{2}(\phi)
=\sqrt{12 V(\sqrt{6}\phi)}\,, 
\label{(5.13)}
\end{equation}
and as a check Eq. (5.12) then takes the form
\begin{equation} 
{1\over V}{dV\over d\phi}=0\,.
\label{(5.14)}
\end{equation}
If the potential (4.18) is assumed, the scalar field being real here 
for simplicity, one finds
\begin{equation}
{1\over V}{dV \over d\phi}={2\over \phi} {(3\phi^{2}-2)
\over (\phi^{2}-1)}\,, 
\label{(5.15)}
\end{equation}
and this does indeed vanish at large $\phi$.

Thus, if we assume that $\phi$ (and hence $\varphi$) 
takes initially a very large value $\phi_{0}={\varphi_{0}\over \sqrt{6}}$, 
we may use (5.11), (5.13) and (5.1) 
to study the first-order system
\begin{equation}
p_{a}={\partial S \over \partial a} ={\partial L \over \partial 
{\dot a}}\,, \; p_{\phi}={\partial S \over
\partial \phi} ={\partial L \over \partial {\dot \phi}}\,, 
\label{(5.16)}
\end{equation}
about which the JWKB wave function is peaked \cite{Espo88}. Its explicit 
form reads as
\begin{equation}
{d\alpha \over dt}=\sqrt{6}\phi^{2} \left(\phi^{2}
-1 \right)^{{1\over 2}}\,, 
\label{(5.17)}
\end{equation}
\begin{equation}
{d\phi \over dt}=-\sqrt{{2\over 3}}{\phi (3\phi^{2}-2) 
\over \left(\phi^{2}-1 \right)^{{1\over 2}}}\,. 
\label{(5.18)}
\end{equation}
Bearing in mind that $\phi$ must be very large in our approximation,
these equations may be rewritten as
\begin{equation}
{d\alpha \over dt}=\sqrt{6}\phi^{2},
\label{(5.19)}
\end{equation}
\begin{equation}
{d\phi \over dt}=-3\sqrt{{2\over 3}}\phi^{2}.
\label{(5.20)}
\end{equation}
It is then immediate to obtain the solution at the end 
of Sec. IV, i.e.
\begin{equation}
\phi={\varphi \over \sqrt{6}}={1\over \sqrt{6}t}, \;
\alpha=-{1\over 12 t^{2}}.
\label{(5.21)}
\end{equation}

\section{Conclusions}

We think that we have illustrated how the introduction of a complex 
scalar field in the inflationary and/or
quintessential scenario can be treated with simple and elegant procedures. 
In our opinion, the main results are:

(1) In our quintessence model, the phase 
of the field must be almost constant,
so that the field may be safely treated as real.

(2) The situation is more involved in the inflationary scenario. It is 
then possible, as already pointed out by
many authors, that the real field is obtained as a limit, by virtue of 
the increasing scale factor present in
the denominator of Eq. (\ref{15}). On the other hand, the introduction
of the perturbing term may introduce a starting point for inflation,
giving rise to some problems of fine tuning. Also the behavior after
the end of inflation can be dramatically changed.

(3) The only possible symmetry of this situation turns out to be the 
trivial one, linked with the cyclic
character of the phase. At this point the Noether symmetry approach has 
proved to be helpful. Even if the potential is
here assumed to be a function of the modulus only, it has to be noted 
that other assumptions are found in the
literature (for example, see Ref. \cite{wei05}, containing a good number 
of references about the possible basic
ways to introduce a complex scalar field in standard cosmology). 
We have shown that this procedure also applies, giving again a cyclic 
phase, in the hessence case.

(4) It is puzzling that the symmetry, which is found for exponential 
potentials in the case of real fields, is
spoiled by the presence of any tiny imaginary component. It is very 
difficult to argue on this particular point because,
as stated in many previous papers, the physical meaning of this kind 
of Noether symmetries (which proved so
fruitful for the solutions of many problems) is still mysterious for us.
Perhaps one can argue that the Noether symmetry only applies to the
background evolution, and has nothing to do with the actual symmetries
of the full theory.

(5) The study of a toy model for the potential in Eq. (4.18),
despite being somewhat ``unphysical'', shows
some intriguing issues, as well as some confirmation of what emerged
in Sec. IV for the dark energy case.

(6) The semiclassical quantum cosmology 
analysis for the scalar-field potential in 
Eq. (\ref{20}) has been obtained for
the first time in the literature in Sec. V. 
The resulting field equations
lead again to the solution at the end of Sec. IV when suitable
approximations are taken into account. This way of supplementing
the results of Sec. IV is important because, as suggested in
Ref. \cite{hawk05}, the evolution of the universe might be semiclassical.

\acknowledgments G. Esposito is grateful to the Dipartimento di 
Scienze Fisiche of Federico II University, Naples, for hospitality 
and support; he dedicates to Maria Gabriella his contribution to
this work.

\end{document}